\documentclass[a4paper]{jpconf}

\usepackage{graphicx}
\usepackage{xcolor}
\usepackage{siunitx}
\usepackage{booktabs}
\usepackage{amsmath,amssymb}
\usepackage{hyperref}
\usepackage[utf8]{inputenc}
\usepackage{microtype}

\usepackage{fancyhdr}
\fancypagestyle{first}{
  \fancyhf{}
  \fancyhead[R]{\textsc{\textcolor{gray}{MCnet-19-18}}}
}
\fancypagestyle{rest}{
  \fancyhf{}
  \fancyhead[R]{\textsc{\thepage}}
}

\begin{document}
\title{Computational challenges for MC event generation}
\author{Andy Buckley}
\address{School of Physics \& Astronomy, University of Glasgow, \textsc{Glasgow}, G12~8QQ, UK}
\ead{andy.buckley@cern.ch}

\thispagestyle{first}
\pagestyle{rest}

\begin{abstract} The sophistication of fully exclusive MC event generation has
  grown at an extraordinary rate since the start of the LHC era, but has been
  mirrored by a similarly extraordinary rise in the CPU cost of state-of-the-art
  MC calculations. The reliance of experimental analyses on these calculations
  raises the disturbing spectre of MC computations being a leading limitation on
  the physics impact of the HL-LHC, with MC trends showing more signs of further
  cost-increases rather than the desired speed-ups. I review the methods and
  bottlenecks in MC computation, and areas where new computing architectures,
  machine-learning methods, and social structures may help to avert calamity.%
\end{abstract}

MC event generation -- i.e.~simulation of the fundamental beam-particle
scattering in a collider experiment, rather than the following interaction of
its decay products with detector systems -- is critical to the the LHC physics
programme. Compared to previous collider analyses, the design and execution of
ATLAS and CMS data analyses in particular have been deeply dependent on
predictions of both background and signal distributions in measured
observables~\cite{Buckley:2011ms}. This greater reliance has also brought
unprecedented demands for precision and accuracy \footnote{The two concepts are
  distinct: precision reflects the formal order of a calculation (and hence
  technical sophistication) and accuracy the resulting uncertainty on
  predictions and the agreement with data.}, with an associated computational
cost.

In this contribution,
derived from a plenary talk at the ACAT~2019 HEP statistics \& computation workshop,
I briefly review the computational bottlenecks in MC
generation for the long-term LHC programme, and a variety of avenues by which
they may be addressed. Many issues in this talk and article may be found
discussed in full detail in the contributions to the 2018 HEP Software
Foundation (HSF) MC workshop:
\url{https://indico.cern.ch/event/751693/timetable/}.

\section{MC generation overview}

As ``MC'' is often used to cover all methods in event simulation using Monte
Carlo methods, let me first clarify that the emphasis here is on fully
differential ``shower + hadronisation generator'' (SHG) codes. In these, a
quantum field theory matrix element (QFT ME) is sampled to provide parton-level
events distributed as the differential cross-section
\begin{equation}
  \mathrm{d}\sigma_\mathrm{hard}(p_1, \ldots , p_n | Q) \sim |\mathcal{M}(p_1, \ldots , p_n)|^2 \, \mathrm{d}\Phi(p_1, \ldots , p_n|Q) \, ,
  \label{eq:dsighard}
\end{equation}
where the $p_i$ are the outgoing leg momenta (we here ignore the coupled
initial-state phase-space term which probes the parton density functions
(PDFs)), $Q$ is the renormalisation scale of the hard-process ME $\mathcal{M}$, and
$\mathrm{d}\Phi$ is the phase-space density for the given configuration. The matrix
element is composed of a finite number of fixed-order terms, expanded in the
strong-force coupling $\alpha_\mathrm{s}$ (and increasingly also in the
electroweak couplings),~i.e.
\begin{equation}
  \label{eq:pertexp}
  \mathrm{d}\sigma_\mathrm{hard} = \mathrm{d}\sigma^\mathrm{LO}_\mathrm{hard}
  + \alpha_\mathrm{s}(Q) \, \mathrm{d}\sigma^\mathrm{NLO}_\mathrm{hard}
  + \alpha^2_\mathrm{s}(Q) \, \mathrm{d}\sigma^\mathrm{NNLO}_\mathrm{hard}
  + \cdots \, .
\end{equation}

The mod-squaring of the matrix element in~eq.~\eqref{eq:dsighard}
mixes amplitudes of different orders in $\alpha_\mathrm{s}$, so that e.g.~ the
$\mathrm{d}\sigma^\mathrm{NLO}$ cross-section correction includes both the
square of a one-emission (``real'') amplitude, and the cross-term composed of
interference between a one-loop (``virtual'') amplitude and the Born amplitude.

This few-body process is then systematically improved toward a realistic event
as would be observed by an ideal collider experiment, by use of various
approximate corrections which distinguish SHGs from ME-only generators. These
most notably include ``parton shower'' (PS) QCD cascades, inclusion of multiple
partonic interactions within the $pp$ collision, and modelling of the
non-perturbative hadronisation process by which coloured partons are resolved
into a collection of primary hadrons. This scheme may be represented formally as
\begin{equation}
  \label{eq:shg}
  \mathrm{d}\sigma \sim \mathrm{d}\sigma_\mathrm{hard}(Q) \times \mathrm{PS}(Q \to \mu) \times \mathrm{Had}(\mu \to \Lambda) \times \ldots \, ,
\end{equation}
where the sequence $Q \to \mu \to \Lambda$ illustrates the evolution down in
energy scale from the ME hard-process/factorisation scale $Q$, through the
renormalisation (etc.) scales of the PS iterations, and eventually cut off by
the dominance of non-perturbative physics at the QCD hadronisation scale,
$\Lambda$.  One of the dominant issues in SHG physics is immediately clear from
eqs.~\eqref{eq:pertexp} and~\eqref{eq:shg}: extra radiation can be generated
both by fixed-order NLO, NNLO, etc. corrections \emph{and} by the PS. The
fixed-order ME computations include full QFT correlations and are important for
describing the few dominant, widely-separated energy flows in collider events,
while the PS ``resums'' large numbers of perturbative QCD emissions in the less
well-separated collinear and soft (low-momentum) phase-space regions where most
emission probability resides. We must be careful not to double-count these
radiative corrections.

The last 20 years have brought a sea-change in the nature of SHG modelling
precision. Back in 2000, when LHC operations were expected to start in around 5
years, the majority of exclusive event generation was still performed using
Born-level matrix elements dressed by parton showers either from the
\textsc{Pythia}~\cite{Sjostrand:2006za} or
\textsc{Herwig}~\cite{Corcella:2000bw} Fortran codes. These showers were both
based on massless $1 \to 2$ parton splitting functions, with corresponding
well-understood deficiencies in phase-space coverage, and the need for an
\textit{ad hoc} ``reshuffling'' scheme to restore Lorentz symmetry to the
post-shower event. The shortcomings of the Born-level matrix element were also
well-understood, and addressed in several cases (notably electroweak boson
production processes) by ``matrix element corrections'' which added the
$V+1~\text{jet}$ matrix element in a consistent way, without double-counting PS
and ME contributions to the same phase-space: these paved the way for further
increases in parton-showered ME precision. Soft-physics modelling was also in a
state far separated from that today, with \textsc{Pythia} having laid the ground
for an eikonal treatment of multiple scattering~\cite{Sjostrand:1987su},
followed by the similar \textsc{Jimmy} model\cite{Butterworth:1996zw} to replace
\textsc{Herwig}'s ageing UA5 parametrisation of the underlying event.

This state of affairs proved sufficient for analyses at HERA and similar
colliders, but Run~2 of the Tevatron drove requirements for higher-precision
matrix elements consistently interfaced to parton showers. With official first
releases around 2002, the \textsc{Alpgen} and
\textsc{MadGraph}~\cite{Mangano:2002ea,Maltoni:2002qb} programs were the
pioneers for ``multi-leg'' tree-level corrections to the Born ME, introducing
the cone-based MLM ME-PS matching scheme, followed by the developments of the
CKKW matching scheme for first the \textsc{Amegic} ME generator and later
\textsc{Sherpa}~\cite{Krauss:2001iv,Gleisberg:2008ta}. The two issues here were
again ones of double-counting phase-space population, both between different
fixed-order processes (``merging'') and between the fixed-order MEs and the
resummed PS (``matching'').

On the same timescale, the MC@NLO program~\cite{Frixione:2003ei} was the
vanguard effort in interfacing one-loop QCD matrix elements (which for full
consistency demand also the tree-level single real emission matrix element) to
parton showers\footnote{Notation is often used loosely in this area: following
  unofficial convention we will here use ``NLO'' to mean matrix elements
  consistently squared at one-loop order, and ``LO multileg'' for tree-level
  corrections beyond the leading (Born) order.}: this saw heavy use for a few
processes, notably top-quark and Higgs physics, but a tight coupling between the
NLO ME subtraction scheme and the PS algorithm delayed the ``NLO explosion'' of
the LHC era until the realisation of the shower-independent \textsc{Powheg}
scheme~\cite{Frixione:2007vw,Alioli:2010xd}.

\section{The rising CPU cost of MC}

The availability of both single-emission NLO SHGs and multiple-emission LO SHGs,
driven by increasing automation of numerical ME calculations, led to a
revolution in experiment expectations of SHG simulation around the LHC
startup. Since 2010, MC generation has gone from being a frequently trivial
element of an experiment's CPU budget to, particularly in the cases of the core
SM $V$+jets and top quark production processes, a substantial consumer in the
range of 15--20\% of 
experiment CPU budget. The driver of this CPU increase has been
the availability of the complex multileg and NLO processes -- by now inevitably
combined into multiple-emission NLO generators e.g~via the MEPS@NLO
formalism~\cite{Nagy:2005aa,Gehrmann:2012yg} implemented in \textsc{Sherpa}, and
the FxFx~\cite{Frederix:2012ps,Alwall:2014hca} one in
\textsc{MadGraph5}~\cite{Alwall:2011uj}, with towers of shower-matched NLO
subprocesses up to some number of final-state particles, followed by even more
LO ME subprocesses. MC is no longer cheap.

This trend is in contrast to the normal direction of travel for CPU budgets at
the LHC experiments because, while the intrinsic complexity of reconstruction
and simulation have increased relatively slowly due to increasing pile-up rates
(ameliorated by algorithmic and implementation improvements), the leaps in
formal precision available as exclusive event simulations come at exponentially
increasing cost in numerical calculation. This has been a particular issue for
the ATLAS experiment which makes heavy use of the \textsc{Sherpa} event
generator: as will be discussed, it is currently more demanding of CPU by
comparison with \textsc{MadGraph5\_aMC@NLO} which dominates CMS simulation
budgets. The single-emission NLO \textsc{Powheg} generator, and a myriad of
other MC codes, are used by both experiments, but with much smaller CPU
consequences -- in this summary we will focus on the most expensive computations.

\begin{figure}
  \centering
  \includegraphics[width=0.48\textwidth]{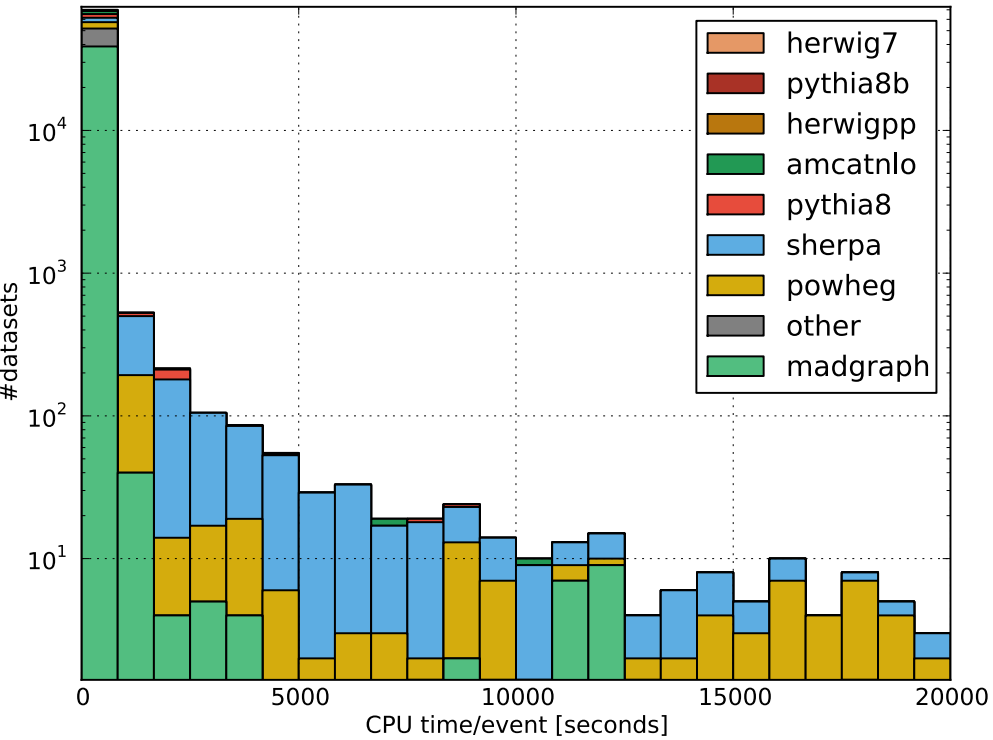}\quad
  \includegraphics[width=0.48\textwidth]{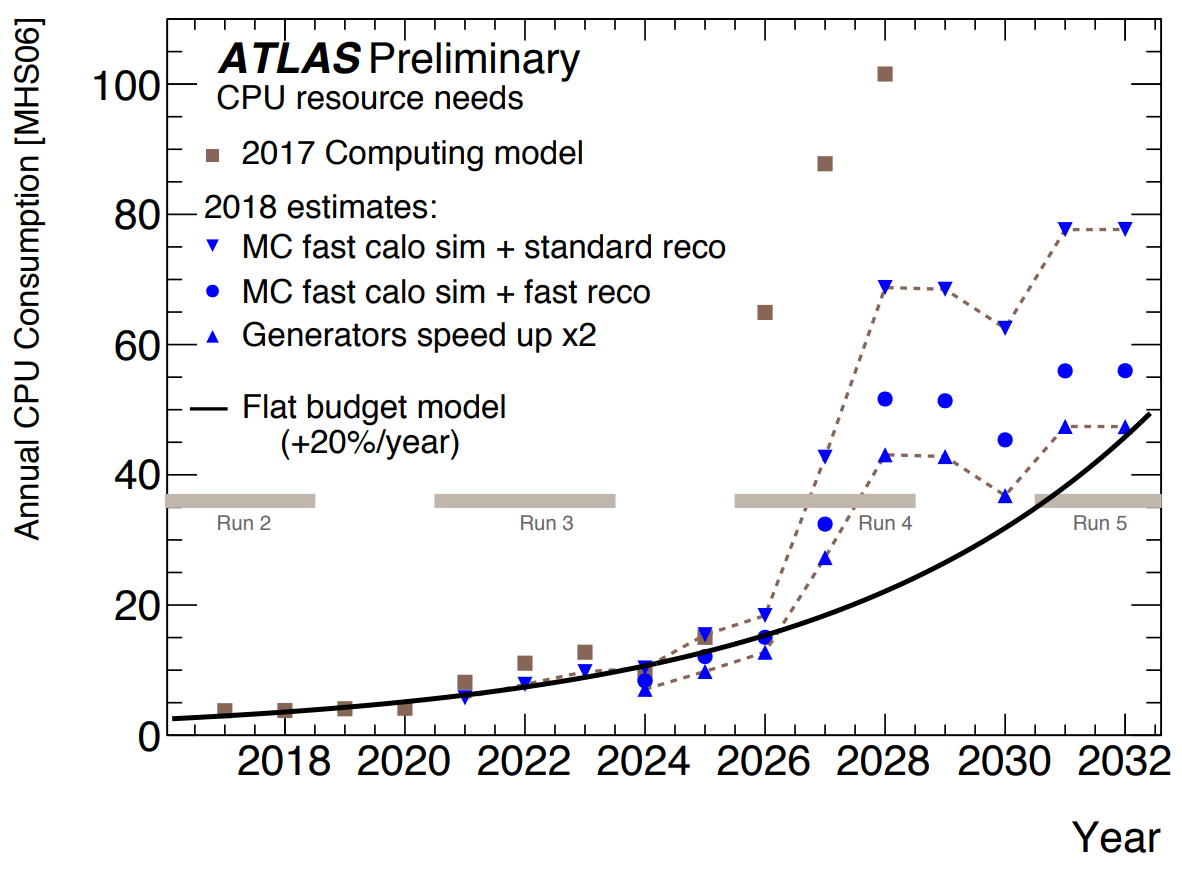}
  \caption{Left: 2015 ATLAS CPU time/event distribution and generator breakdown,
    plot from Josh McFayden. Right: ATLAS CPU demand vs. availability projection
    to HL-LHC operation.}
  \label{fig:atlascpu}
\end{figure}

As shown in Figure~\ref{fig:atlascpu}, this trend cannot continue: the CPU
requirements of the High-Luminosity LHC programme are already insufficient for
Grid-based MC generation alone, even with optimistic assumptions about evolution
of computational purchase power and the ability to obtain algorithmic
speed-ups. A further step change in formal precision, e.g.~from 1-loop NLO to
2-loop NNLO as the core-process standard for SM SHG generation, may come at such
unacceptable cost that it is of academic interest only, priced out of the
straitened WLCG budgets by a further exponential step in cost per event.

It is clearly unthinkable that the vast public investment, and the scientific
and engineering achievements, of the long-term LHC programme be limited in
physics impact by an inability to generate sufficient MC event statistics at the
required precision. Something needs to change: in the rest of this article I
will summarise the most promising candidates.

\section{Challenges and innovations in matrix element phase-space integration}

At the core of MC generation, and the most expensive part for high-multiplicity
processes, is the integration -- synonymous with asymptotic sampling -- of the
(squared) QFT matrix element over the allowed phase-space of incoming and outgoing legs'
momentum configurations, cf.~eq.~\eqref{eq:dsighard}.

This is intrinsically a complex problem because the squared matrix elements are
extremely ``peaky'' functions: the majority of their probability density is
located in the vicinity of kinematic divergences, and na\"ive ``flat'' sampling
of the phase-space will produce abominable statistical convergence. Even with
improved sampling proposals, such issues are seen in the form of a wide
distribution of sample \emph{weights} or equivalently a poor efficiency for
generation of unweighted events -- both major problems for LHC consumers of
precision MC generators.

\subsection{Sampling and event weights}

Efficient MC generation requires efficient sampling of the peaky ME function
over the partonic phase space. The ideal sampling would be to draw the samples
themselves from the asymptotic ME distribution, i.e. knowing the answer before
we began! In practice, of course, this is not known \textit{a priori} and so
approximate proposal distributions are needed, with standard sampling techniques
used to recover the asymptotic distribution by discarding or re-weighting the
samples from the approximate one. The aggregation of these imperfect phase-space
mappings is the major cause of poor efficiencies in MC event sampling at high
fixed orders in $\alpha_\mathrm{s}$.

\begin{figure}
  \centering
  \includegraphics[width=0.48\textwidth]{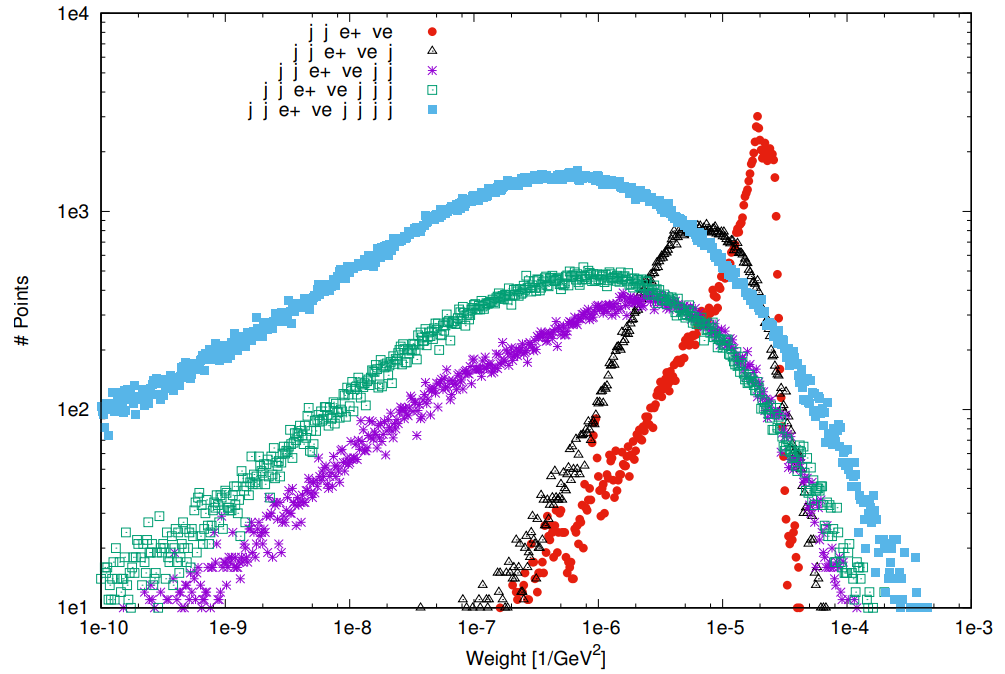}\quad
  \includegraphics[width=0.48\textwidth]{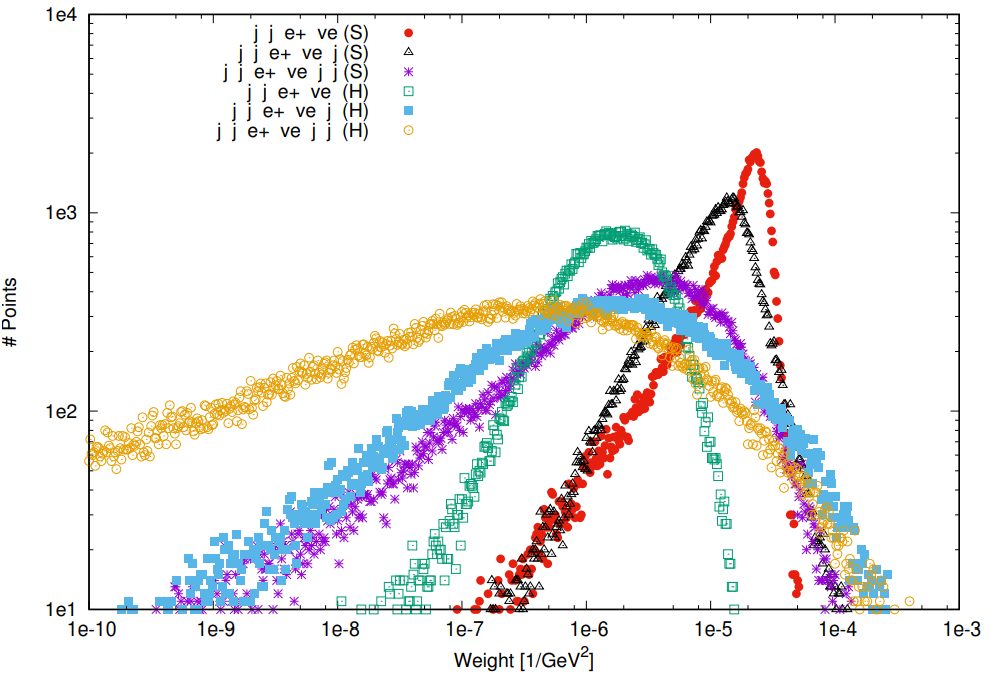}
  \caption{Sherpa event weight distributions for LO (left) and NLO (right,
    decomposed as MC@NLO soft/hard event classes) for different
    $Wjj + n\text{-jets}$ ME multiplicities. The broadening of distributions
    toward lower weights with higher multiplicity, particularly at NLO, reflects
    the increasing difficulty of approximating the ME with current sampling
    proposal distributions. Plots from Marek Schoenherr.}
  \label{fig:weightdbns}
\end{figure}

The ideal, maximally efficient proposal distribution would by definition always
have unit weights (the proposal density exactly matches the real one) but in
practice the sample weights have a tail to lower values (the proposal included
phase-space points which were uninteresting). Example weight distributions for
\textsc{Sherpa} multi-leg $W+\text{jets}$ event sampling are shown in
Figure~\ref{fig:weightdbns}. Greater than unit weights are an even worse problem
in principle: the proposal density underestimated the maximum, probably due to a
failure in pre-sampling. Both problems feed into observables computed from the
sampled events, as poor statistical convergence and as single-event spikes
respectively.

In practice weighted events are often difficult to use for experimental purposes
because, unlike for most phenomenology purposes, the bulk (or at least a
comparable amount) of their CPU cost is still to come in the form of detector
simulation. It is usually a better strategy to \emph{unweight} events to obtain
a (still unbiased) sample of high-weight events than to spend processing power
expensively running Geant\,4 for events with unrepresentative phase-space
configurations. This sample rejection from broad distributions of weights leads
to a further 
inefficiency which, in combination with already CPU-intensive computation of the
ME value for each sample, explains the rocketing CPU cost of state-of-the-art
SHG MC event samples. Some current processes can even take
$\mathcal{O}(24~\text{hours})$ of CPU time per event!

\subsection{Sampling improvements}

The now-standard approach to phase-space integrals is transformation of the
integral to reflect the na\"ive singularity structure of the cross-section, which
typically diverges for soft and collinear configurations. This pattern continues
into higher-order LO phase-space, with probability density increasing as legs'
momenta become either very small, or parallel to another leg. The
``multi-channel'' decomposition of the integral into the weighted sum of its
pole structures was a key step in automation of matrix element
calculations~\cite{Maltoni:2002qb}, but its imperfections become evident at high
multiplicities (where the combinatorics of divergences are formidable), and due
to further modifications of the divergences by loop amplitudes.

\begin{figure}
  \centering
  \raisebox{1.3em}{\includegraphics[width=0.36\textwidth]{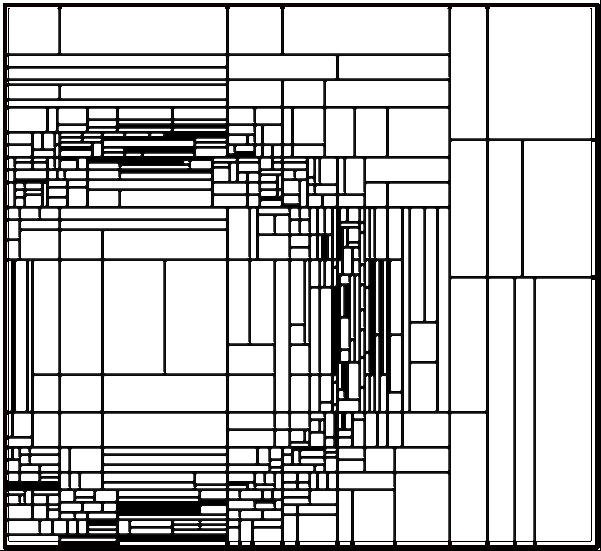}}
  \quad
  \includegraphics[width=0.55\textwidth]{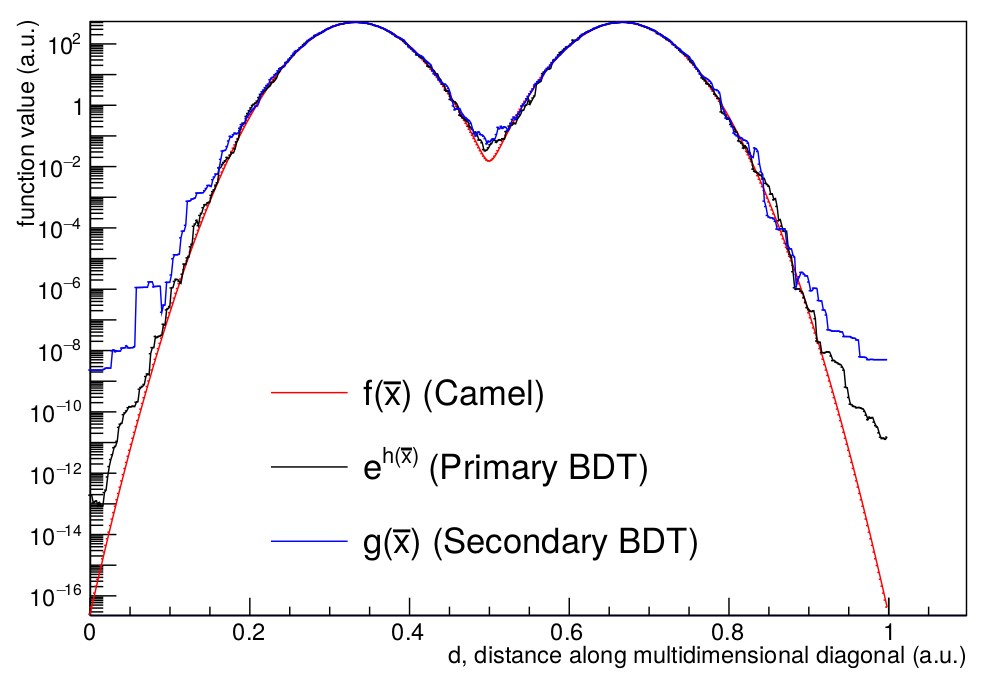}
  \caption{Left: an illustration of 2D \textsc{Foam} for improved sampling of a
    non-factorisable 2D ring distribution. Right: convergence of the
    multidimensional ``Camel'' function using the BDT-integration method
    (deviations from optimal visible on tails and between the humps).}
  \label{fig:foambendavid}
\end{figure}

A particular limitation here is the widespread use of the \textsc{Vegas}
adaptive integration/sampling algorithm for ME/PS
integration~\cite{PETERLEPAGE1978192}. This allocates dynamically resized bins
in each integration parameter independently, with many smaller bins allocated to
regions of (dynamically discovered) probability density. \textsc{Vegas} gives
good performance for separable distributions (provided the factorization is
identified in advance), but is unable to cope efficiently with non-factorisable
integrands. This issue has been realised for some time, but has not yet received
mainstream attention in MC codes -- perhaps for reasons to be discussed in
Section~\ref{sec:lowfruit}. The obvious step up in sophistication is the
multidimensional adaptive binning scheme of \textsc{Foam}~\cite{Jadach:2002kn},
which uses nested hypercubes in place of single-dimensional bins,
cf.~Figure~\ref{fig:foambendavid} and hence has greater purchase on
moderate-dimensional non-factorisable integrands. But still the scaling of
cross-section integration is demanding.

An exciting prototype study~\cite{Bendavid:2017zhk} has applied machine learning
(ML) to this problem, built on the identification of first boosted decision
trees (BDTs) and latterly deep neural networks (DNNs) as generalisations of
hypercube binning. This approach has not yet been applied in anger to MC
integration, as significant technical work is required to rework the integration
machinery of a real-world generator code, but the preliminary study's use of a
standard ``Camel'' awkward function in 4 and 9 parameter dimensions produced
remarkable gains in variance reduction for fixed sample number, relative to both
VEGAS and \textsc{Foam}, as shown in Table~\ref{tab:bendavid}. This pathfinding
work raises the possibility of ML-assisted integration coming to the rescue in
the battle of MC precision vs. CPU budget.

\begin{table}
  \centering
  \begin{tabular}{lrr}
    \toprule
    Algorithm & Number of evaluations & $\sigma_w/\langle w \rangle$\\
    \midrule
    4D Camel function & & \\
    \textsc{Vegas}          & 300k &  2.820 \\
    \textsc{Foam}           & 3.8M &  0.319 \\
    Generative BDT (staged) & 300k &  0.077 \\
    Generative DNN (staged) & 294k &  0.030 \\
    \midrule
    9D Camel function & & \\
    \textsc{Vegas}          & 1.5M & 19.000 \\
    \textsc{Foam}           & --~~ &  --~~ \\
    Generative BDT (staged) & 3.2M &  0.310 \\
    Generative DNN (staged) & 294k &  0.081 \\
    \bottomrule
  \end{tabular}
  \caption{Performance of the best BDT and DNN models for integrator
    improvement, compared to \textsc{Vegas} and \textsc{Foam} (where feasible)
    on 4- and 9-dimensional Camel test functions.}
  \label{tab:bendavid}
\end{table}

A study with similar motivations was performed in a Masters
thesis~\cite{KrauseThesis}, this time employing several ML techniques in an
attempt to specifically learn the integration phase-space of the $Z+q\bar{q}gg$
process. This, again preliminary, study illustrates the importance of domain
knowledge to application of ML machinery, with a systematic study of three
distinct bases in event-kinematics features, exploiting knowledge of Lorentz
invariance and other physical principles in attempts to find the optimum
parametrisation. Again there were indications of significant gains in
performance, with a factor of 750 speed-up between sampling from the ML model as
compared to standard multi-channel sampling of the full matrix element\dots but
at the cost of a $\times 3$ mis-modelling of the weight distribution. This is
too inaccurate to use the ML itself as a fast MC generator, but compared to the
weight distributions in Figure~\ref{fig:weightdbns} it seems again that ML
techniques may have important future roles to play in efficient generation of
ME-integrator proposal densities.


\subsection{NLO subtraction and negative weights}

In addition to the problem of broad weight distributions, and the resulting
unweighting inefficiencies, the consistent treatment of loop amplitudes in NLO
MC generation additionally introduces problematic rates of \emph{negative
  weights}.

These arise because both the real and virtual $\mathrm{d}\sigma$ corrections
have infrared divergences, in the soft and collinear limits of the real and loop
momenta. In a tree-level calculation, the fixed-order real divergence is
typically hidden by a phase-space cut, with the intention that approximate
resummation of multiple emissions via a parton shower will consistently fill the
divergent phase-space (this is the role of ``LO matching''), but full NLO
cross-section normalisation requires consistent combination of both fixed-order
terms (as well as the PS again -- ``NLO matching''). Following the
Block-Nordsieck, YFS, and KLN
theorems~\cite{bloch1937note,yennie1961infrared,kinoshita1962mass,lee1964degenerate},
in the indistinguishable limit of an unresolveable real emission, the two
opposing fixed-order singularities cancel, leaving a finite residual. As
infinities cannot be directly worked with in numerical codes, this is
canonically implemented via \emph{subtraction} of the opposing divergence
structure $\mathrm{d}\sigma_\mathrm{S}$ from both the real and virtual terms,
making them independently finite:
\begin{align}
  \label{eq:subtraction}
  \mathrm{d}\sigma &= \mathrm{d}\sigma_\mathrm{B} + \mathrm{d}\sigma_\mathrm{NLO}\\
                   &= \mathrm{d}\sigma_\mathrm{B} + (\mathrm{d}\sigma_\mathrm{V} + \int \! \mathrm{d}\Phi_1 \, \mathrm{d}\sigma_\mathrm{S}) +  \left( \mathrm{d}\sigma_\mathrm{R} - \mathrm{d}\sigma_\mathrm{S} \right) \, .
\end{align}

Of course, these singular phase-space regions are the same ones whose difficulty
in mapping led to ME integration inefficiencies, and the same problem -- only
more-so -- plagues NLO calculations, with the new issue that insufficient sample
densities can produce negative cross-section estimates via the subtraction
scheme. Negative weights essentially count double toward generation
inefficiencies because not only do they not add statistical convergence but they
actively subtract it, cf. the variance formula
$\sigma^2 = \langle \sum w^2 \rangle - \langle \sum w \rangle^2$ in which the
first term always increases but the second will be reduced by a mix of positive
and negative weights. The MC@NLO subtraction formalism (which includes PS
splittings in the subtraction) naturally induces negative-weight fractions of
around 25\%, corresponding to an effective halving of the statistical power of
an event sample.

The effect of higher complexity and negative weights at NLO were illustrated by
Valentin Hirschi to this workshop, in his demonstration of generation CPU for
addition of 1 and 2 extra gluons to the $d\bar{d} \to ZZ$ process: at LO this
scales rapidly from \SI{7}{\micro\second} to \SI{35}{\micro\second} to
\SI{220}{\micro\second}, and are respectively made slower by factors of $10^2$,
$10^3$, and $10^4$ at one-loop order.

In the absence of a revolutionary new formalism for NLO
calculations, the best hope is again that ML or other improvements in
phase-space mapping will reduce the incidence of negative weights\dots or, as
will be discussed in Section~\ref{sec:lowfruit}, to perhaps entirely reconsider
the LHC experimental demand for SHG NLO events in favour of less sophisticated,
but pragmatic, approximations.


\section{Matrix-element merging}

So far we have focused on ME phase-space integration as a leading bottleneck for
high-multiplicity process generation. But in particular for the \textsc{Sherpa} event
generator, the combination of samples from different-multiplicity
matrix-element terms also adds a very significant computational cost. The
concept behind merging is again (as is almost everything in pQCD MC) to use the
highest-accuracy calculations in appropriate phase-space regions -- i.e.~the
most sophisticated matrix-elements possible for configurations with hard, widely
separated partons, and less sophisticated ones enhanced by the PS in soft and
collinear regions where multiple emissions dominate -- and to avoid
double-counting in the process.

\begin{figure}
  \centering
  \includegraphics[width=0.65\textwidth]{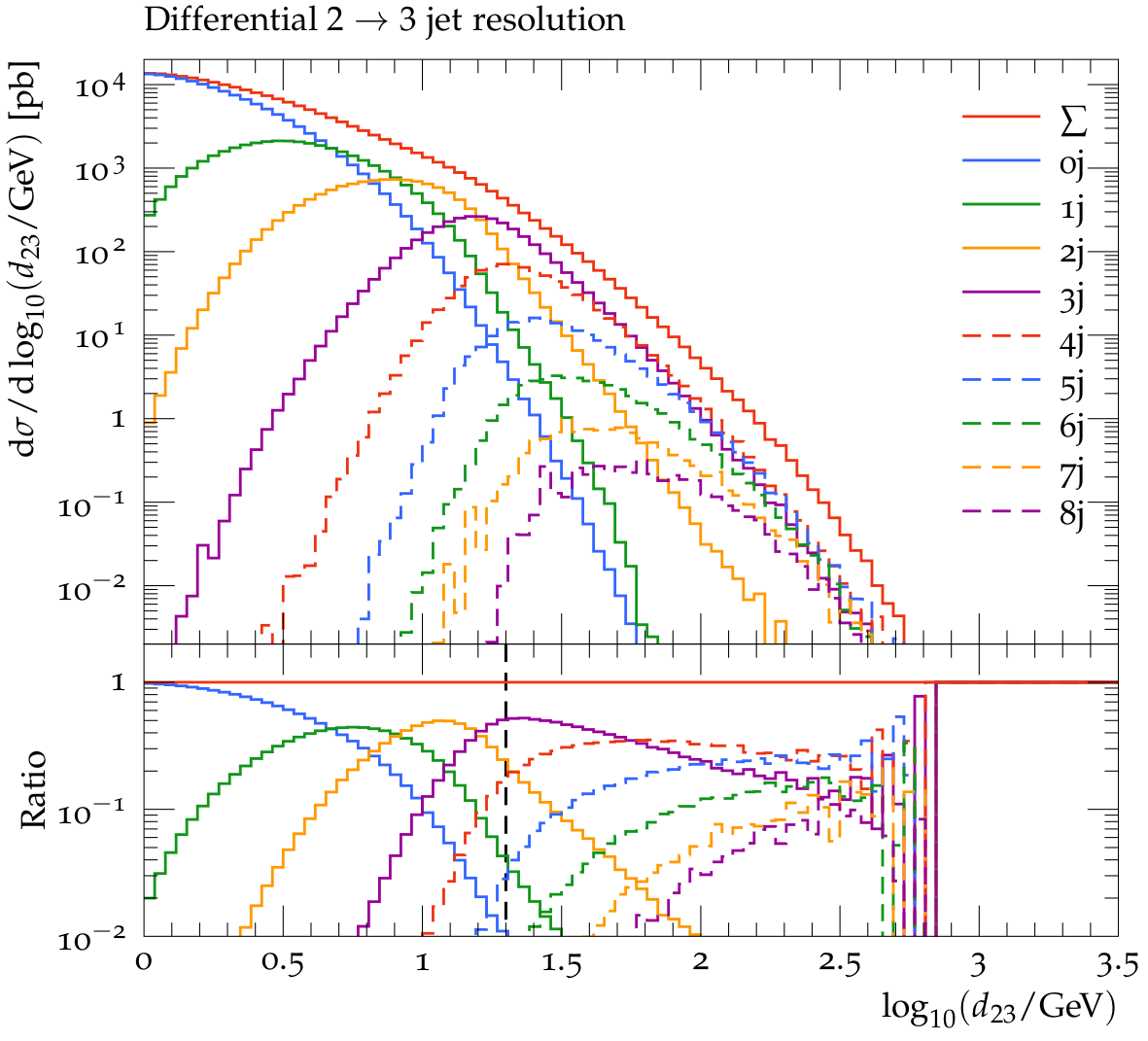}
  \caption{Illustration of ME merging in Sherpa, showing the contributions of up
    to 8 additional ME partons above the Born level, summing to an inclusive
    distribution. The smooth suppressions of the higher parton multiplicities at
    low splitting scales $d_{23}$ are introduced by the CKKW phase-space slicing.}
  \label{fig:mergingcontribs}
\end{figure}

\begin{figure}
  \centering
  \includegraphics[width=0.85\textwidth]{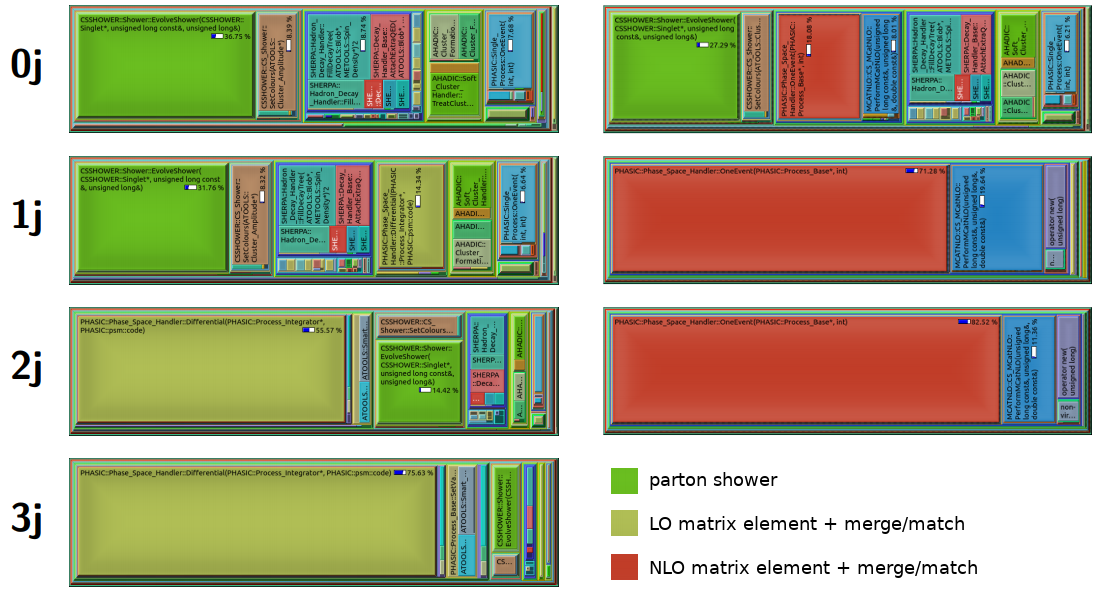}
  \caption{Valgrind CPU maps of Sherpa $W+n\text{-jet}$ event generation for LO
    (left) and NLO (right) event generation. The rapid dominance of ME sampling
    and merging with increasing multiplicity, especially at NLO, is clearly
    visible. Plots from Marek Schoenherr.}
  \label{fig:sherpaperf}
\end{figure}

This is achieved by slicing the phase-space into orthogonal regions in which
each kind of effect dominates -- an unphysical distinction, but a necessary one
in a world of finite intellectual and computational resources. Since QFT
singularities again drive the relative dominance of fixed-order vs resummation
effects, the CKKW merging procedure in particular uses a modification of the
$k_t$ jet clustering scheme (adapted to be aware of \textsc{Sherpa}'s PS
splitting functions) to delimit these phase-space slices, illustrated by
Figure~\ref{fig:mergingcontribs}. But this algorithm is itself CPU-expensive!
The rapid CPU scaling of \textsc{Sherpa}'s components in LO and NLO $W$
production, with increasing numbers of ME jets, is shown in
Figure~\ref{fig:sherpaperf}.

In fact, direct process-to-process comparisons of \textsc{Sherpa} to \textsc{MadGraph5} indicate a
factor of $\times 4$ performance difference, which can be reduced by more than a
factor of $2$ simply by using a simple scale-based calculation in place of the
formally more accurate CKKW clustering, with minimal observable effect. This
process-specific approximation, and similar insights for other bulk SM
background samples used by the LHC experiments, is the basis for much of ATLAS'
hoped-for MC speed-up by a factor of two in the HL-LHC era: not all MC
improvements must come from brute-force calculation, but also judicious use of
approximations.

\section{Machine learning in ME/PS matching}

Judicious approximations bring us to a last consideration in the physics of
state-of-the-art SHG generation. I wish to highlight for once not a performance
concern, but an example of how ML techniques may allow ``missing'' physics to be
filled in without explicit calculation.

The same issues of double-counting and of using jet clustering to obtain
formally more precise scale estimations are found in not just ME merging but
also the matching of fixed-order MEs to multiple-emission parton showers. The
\textsc{Minlo} (multi-scale improved NLO) method is an approach to improve the \textsc{Powheg} NLO
matching scheme by discovery of an event-specific ME emission scale and matching
the fixed-order real emission to the Sudakov form factor for resummed QCD
emissions at the same scale.

In Reference~\cite{Carrazza:2018mix}, this approach is taken in the matching of the NLO
single-top + jet (STJ) process to a parton shower. The \textsc{Minlo} matching
cross-section is given by
\begin{equation}
  \label{eq:minlo1}
  \mathrm{d}\sigma_\mathcal{M} = \Delta(y_{12}) \left[
    \mathrm{d}\sigma^\mathrm{STJ}_\mathrm{NLO} -
    \Delta(y_{12})|_{\alpha_\mathrm{s}}  \mathrm{d}\sigma^\mathrm{STJ}_\mathrm{LO}
  \right] \, ,
\end{equation}
where $\mathrm{d}\sigma^\mathrm{STJ}_\mathrm{LO/NLO}$ are the LO and NLO STJ
cross-sections, and $\Delta(y_{12})$ and $\Delta(y_{12})|_{\alpha_\mathrm{s}}$
are respectively the Sudakov form factor and its $\alpha_\mathrm{s}$ term,
evaluated at the $k_t$-clustering scale $\sqrt{y_{12}}$. As intended, this
method replaces the fixed-order divergence of the STJ cross-section (as the
extra jet becomes unresolved) with a smooth Sudakov suppression, giving STJ
predictions accurate at NLO+NLL (next-to-leading logarithm resummation). But
there is a remaining defect: this unresolved-jet limit can be understood as the
ST limit of the STJ process, but the \textsc{Minlo} construction is not
automatically accurate at NLO+NLL for ST. To do so, resummation terms at NNLL
would need to be included in the Sudakov factors $\Delta(y_{12})$ in
eq.~\eqref{eq:minlo1}. Calculation of such terms is in general a major exercise,
which has not yet been automated to the same extent as fixed-order NLO
calculations.

\begin{figure}
  \centering
  \begin{minipage}[b]{0.41\linewidth}
    \includegraphics[width=\textwidth]{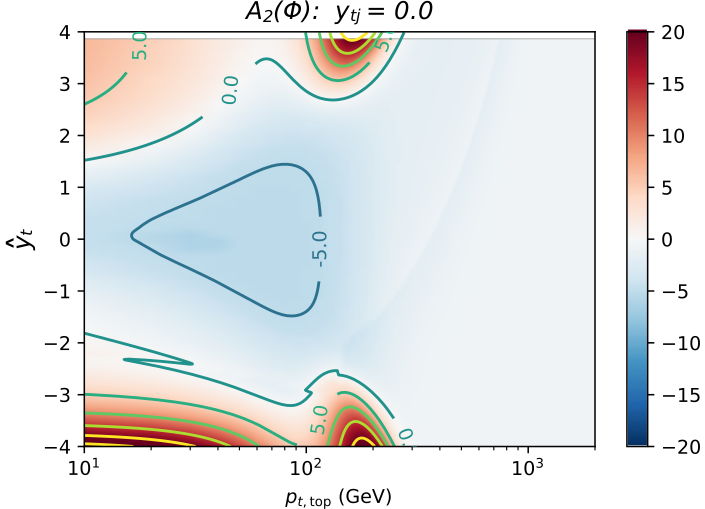}\\
    \includegraphics[width=\textwidth]{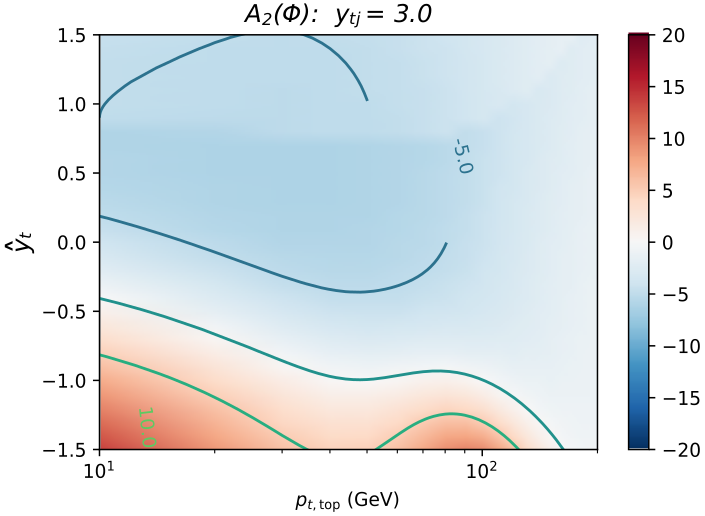}
  \end{minipage}
  \quad
  \raisebox{0em}{\includegraphics[width=0.51\textwidth]{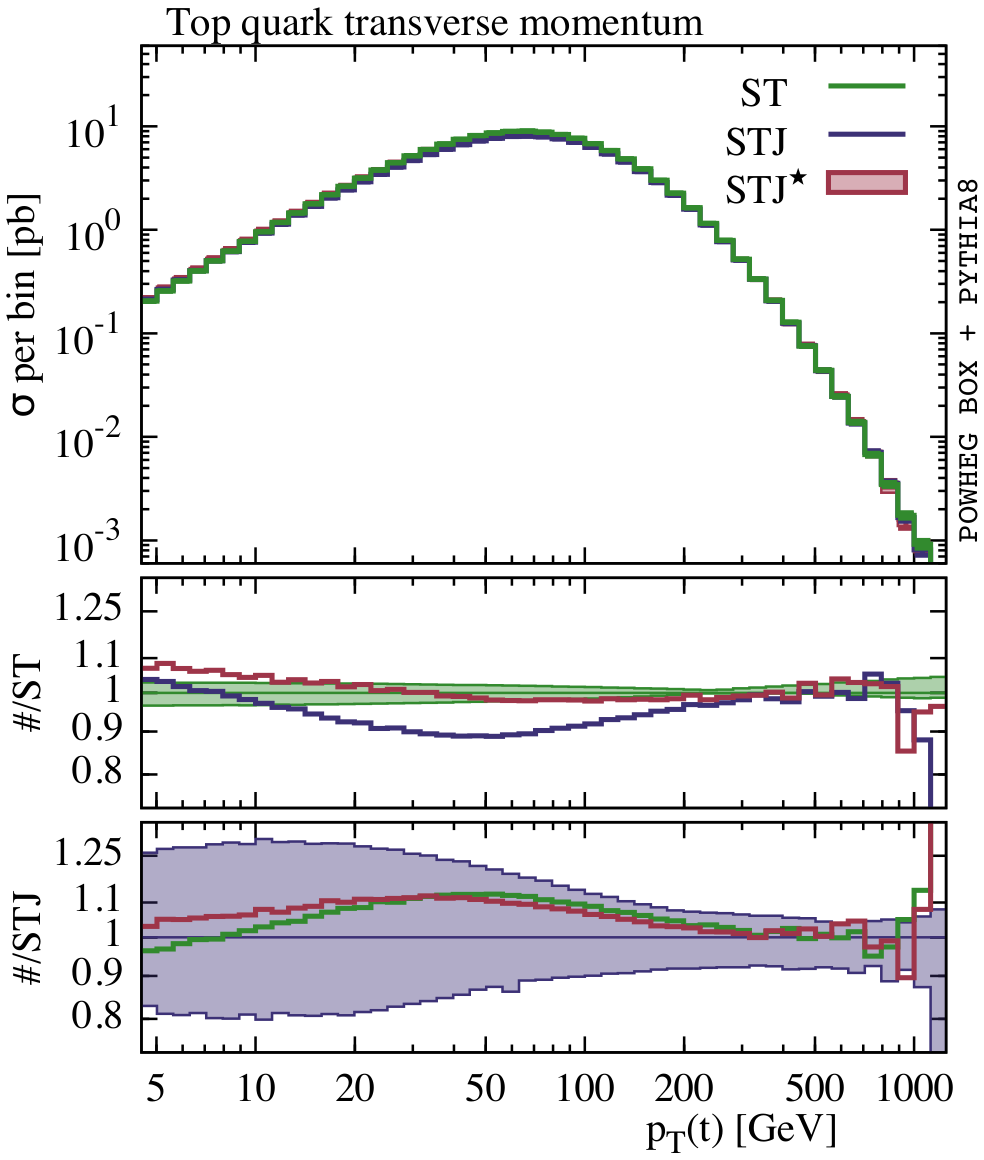}}
  \caption{Left: a slice of the ML-approximated $\mathcal{A}_2(\Phi)$
  function. Right: top-quark transverse momentum, comparing standard
  \textsc{Minlo} ST and STJ results with the ML-improved ``STJ*''
  calculation. STJ* is, as intended, simultaneously consistent with both the ST
  and STJ processes.}
  \label{fig:imprminlo}
\end{figure}

This is where machine learning comes in: ML techniques can in principle
construct the missing Sudakov function by demanding boundary-condition
consistency between the STJ and ST process limits. To achieve this the authors
of Ref.~\cite{Carrazza:2018mix} propose an ``STJ*'' ansatz of modifying the
exponent of $\Delta(y_{12})$ by an $\mathcal{O}(1)$ function of the Born
kinematics, $\mathcal{A}_2(\Phi)$,
\begin{equation}
  \ln \delta \Delta(y_{12}) = -2 \int_{y_{12}}^{Q^2_{bt}}
  \frac{\mathrm{d}q^2}{q^2} \left( \frac{\alpha_\mathrm{s}}{2\pi} \right)^2 
  \mathcal{A}_2(\Phi) \, \ln \! \frac{Q^2_{bt}}{q^2} \, ,
\end{equation}
such that the ST differential cross-section is given by
\begin{equation}
  \frac{\mathrm{d}\sigma^\mathrm{ST}_\mathrm{NLO}}{\mathrm{d}\Phi} =
  \int \mathrm{d}y_{12} \, \frac{\mathrm{d}\sigma_{\mathcal{M}}}{\mathrm{d}\Phi \, \mathrm{d}y_{12}} \,
  \delta \Delta(y_{12}) \, .
\end{equation}
The estimation of $\mathcal{A}_2(\Phi)$ then corresponds to a deconvolution of
the STJ cross-section with $\delta \Delta(y_{12})$ over the Born phase space,
and was performed using an optimised neural net architecture with loss function
\begin{equation}
  \mathcal{L} = \sum_i^{N_\mathrm{bins}} \left[
    \sum_j^{N_\mathrm{evt}} w^\mathrm{ST}_{i,j} -
    \sum_k^{N'_\mathrm{evt}} w^\mathrm{STJ}_{i,k} \mathrm{e}^{\tilde{\mathcal{A}}_2(\Phi_i) \mathcal{G}_2}
  \right]
\end{equation}
where the $i$-sum is over discrete bins in the phase-space $\Phi$, the $j$- and
$k$-sums are over events in ST and STJ event samples,
$\tilde{\mathcal{A}}_2(\Phi_i)$ is the trial function, and $\mathcal{G}_2$ is a
$\Phi$-independent term from NNLL resummation. A slice of the
$\mathcal{A}_2(\Phi)$ function and the resulting improved-\textsc{Minlo} STJ
prediction of top-quark $p_T$ are shown in Figure~\ref{fig:imprminlo}. Such uses
of new non-parametric methods for improvement of matched NLO calculations are an
exciting new direction which we may expect to see more of in future.

\section{New computing architectures}

Computing architectures have long been of subleading interest to SHG MC
developers, since the event sampling is embarrassingly parallel: just use
distinct random-number generator seeds for each run, and linear scaling is
trivial. But the rise of high-multiplicity matrix element calculations, in
particular at NLO, has changed the game: large memory requirements require
memory sharing between logical cores, and complex phase-space sampling demands
\emph{coordinated} adaptive sampling between many compute nodes. In addition,
the availability of architecture developments like vectorization (in CPU, GPU \&
tensor processing) may also offer benefits that cannot responsibly be ignored
but whose exploitation will require significant technical effort.

The first trend being driven by rising computational costs is deconstruction of
the neat black-box operation of the C++-era SHG
codes~\cite{Sjostrand:2007gs,Bahr:2008pv,Gleisberg:2008ta}. \textsc{Sherpa} in
particular is a sophisticated combination of a matrix-element assembler,
sampler, matching \& merging algorithm, parton showers, electroweak corrections,
MPI, and hadron-decay modelling, all in one package and run through one
command. The huge cost of ME integration long ago led to a decoupled initial
scan mode to map the phase space for efficient use in event generation (the
so-called ``gridpack'' mode, following \textsc{MadGraph5\_aMC@NLO}
nomenclature), but \textsc{Sherpa} still attempts to handle all $n$-leg ME
generation, merging, matching, showering, etc. inside a single generation run --
with the result that it runs as fast as its slowest sub-process. This, in
addition to the CPU cost of CKKW clustering discussed earlier, largely explains
the high CPU bill of \textsc{Sherpa}-heavy ATLAS as compared to
\textsc{MadGraph5}-dominated CMS.

\begin{figure}
  \centering
  \includegraphics[width=0.48\textwidth]{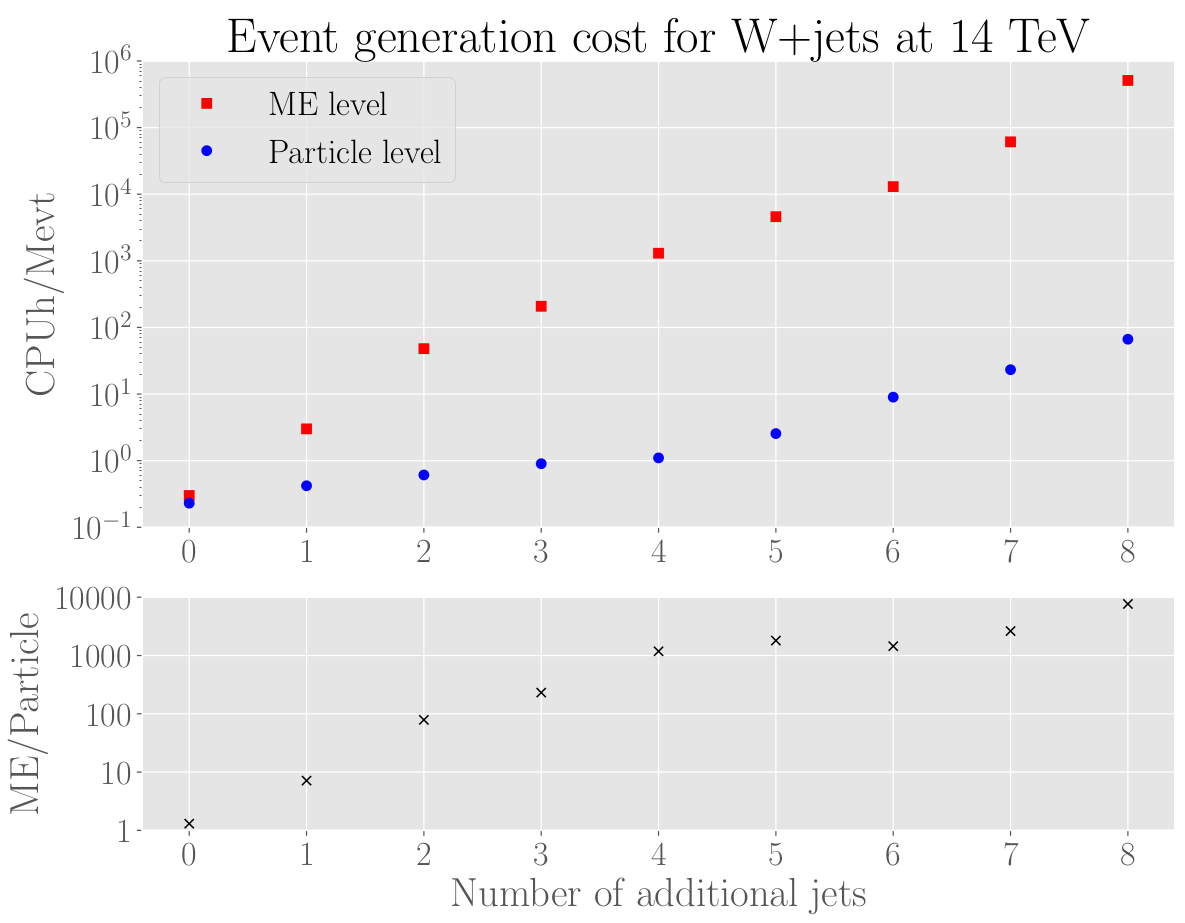}\quad
  \includegraphics[width=0.48\textwidth]{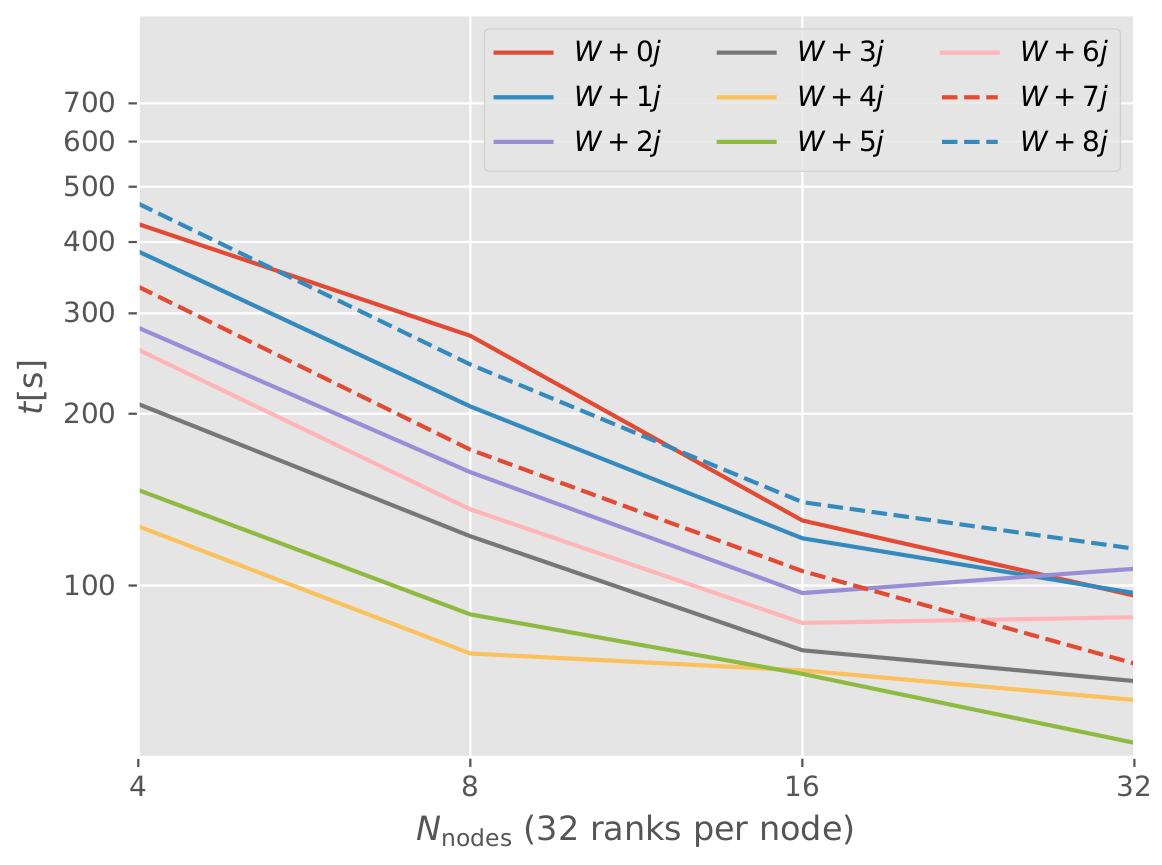}
  \caption{Left: relative scaling of ME-level and Particle-level (rest of SHG)
    CPU cost with ME sub-process multiplicity: ME is exponentially more expensive
    at high multiplicity, driving use of HPC for high-multiplicity ME
    sampling. Right: time-scaling of ME integrations with the degree of MPI
    parallelism. Plots from Holger Schulz.}
  \label{fig:sherpahpc}
\end{figure}

This deficiency, and the increasing availability of high-performance computing
(HPC) facilities with multi-node coordination via the message passing interface
(MPI), has driven a project to factorise \textsc{Sherpa}'s generation strategy so that
the number-crunching capacity of the big facilities is used to generate each
parton-multiplicity ME event sample separately, for later merging and downstream
physics-processing. This has involved developing a new, partonic event format
based on the HDF5 technology, which supports parallel read and write, as opposed
to the write-locked plain text LHE format which is the current
standard~\cite{Alwall:2006yp}. As shown in Figure~\ref{fig:sherpahpc}, the CPU scaling of
ME generation is exponential in parton multiplicity, while the shower \&
matching elements scale more respectably: we may hence be entering Run~3 of the
LHC with a new hybrid strategy for precision event generation, in which
high-multiplicity partonic event samples are generated on HPC resources, and
lower-multiplicities and downstream generator components run on WLCG Grid
resources as now. The potential is very high: Figure~\ref{fig:sherpahpc}
illustrates the HPC scaling of various $W+n~\text{jets}$ MEs, up to an
extraordinary $n = 8$: with this system, generating and analysing 100M events at
up to 8-jet order has been performed in 25 minutes!

A final mention of GPU technology is worthwhile, even though this technology has
not yet delivered more than prototype demonstrators in event
generation. \textsc{MadGraph5} investigations of LO ME sampling with GPUs began in
2009\footnote{NLO generation on GPUs is currently unfeasible since the large
  memory size of NLO matrix elements exceeds that available on GPU devices.},
and all SM processes have been supported in that mode since 2013, but there is
little production use so far. It remains to be seen whether GPU or conventional
HPC resources offer more potential for LHC simulation campaigns: effort will
naturally follow available facilities to some degree. An intriguing angle is the
increasing sophistication and ``auto-GPUing'' of vector processing tasks via
machine-learning frameworks such as \textsc{TensorFlow}: these could remove the language
barrier that discourages MC authors from focusing on uncertain uptake of GPU
processing, by having the code automatically generated. This would also be an
appealing route to incorporation of ML-assisted integration strategies. The
ML/data-science world being heavily oriented towards Python, it seems that
\textsc{MadGraph5} is particularly well-positioned to explore and benefit from these
new technologies.





\section{Low-hanging fruit?}
\label{sec:lowfruit}

Our final duty is to step back briefly from the technicalities, both
mathematical and technological, of high-precision MC generation, and consider
whether there are simpler strategies that might also achieve LHC aims.

There is an increasing realisation that NLO precision may not always give better
accuracy and that sometimes the optimal solution for experimental purposes is
both more approximate and much cheaper computationally. To understand this, we
must appreciate that SHG MC events are used in two ways by collider experiments:
\begin{enumerate}
\item for tests of state-of-the-art theory against experiment data;
\item as the best available emulation of how events are distributed and reconstructed.
\end{enumerate}
The latter is in fact more common: when estimating the backgrounds to new
physics searches, or deriving the detector response for use in observable
unfolding, a high formal precision gives confidence in a simulated event sample,
but we would also be happy with any large, accurate event sample that fell out
of the sky, regardless of origin. In practice, even the highest-precision MC
often gets re-weighted either differentially or in normalisation to better match
data in control regions. This reality in which formal precision is not something
essential and sacrosanct -- except of course when explicitly testing theory --
offers us a strategic viewpoint different to that assumed thus far: can we
achieve experimental accuracy via less formally satisfying (and expensive)
ansaetz?

Given, in particular, the large per-event cost premium of high-multiplicity NLO
over its tree-level LO equivalent, this is not a stupid question: there may be
more physics value in using the available CPU for higher event statistics than
for higher-$n$ $\text{N}^n\text{LO}$ and $\text{N}^n\text{LL}$ computations. It
is also useful to note that the headline achievement of NLO-matched SHG
simulation is not actually used for most major SM samples: the total
cross-section, so carefully preserved by matching algorithms, is frequently
obviated by normalising the SHG sample to a state-of-the-art parton-level
calculation at e.g.~NNLO+NNLL order. Indeed, as differential calculations at
such orders and in directions such as beyond-leading-colour PS become available,
such re-weightings are also becoming differential -- with great care. Making
multidimensional re-weightings of this type efficient is again an area where ML
methods may have a role to play: when approximations can add value, ML is our
best hope of systematically making better approximations. NLO matching is not
rendered pointless by this strategy, since the preservation of NLO-accurate
observables\footnote{Note that not all observables from an ``NLO''-branded MC
  generator are NLO-accurate. The classic counterexample is the boson (or
  leading-jet) $p_T$ in inclusive Higgs or vector boson production: the LO
  process gives a trivial zero-$p_T$, and it is this $2 \to 1$ amplitude that is
  stabilised by the addition of the NLO loop term; at ``NLO'' the boson $p_T$
  becomes non-trivial, but only at LO accuracy due to the unstabilised
  real-emission term. Further observables such as the second-jet $p_T$, will
  only be accurate to LL since they don't even exist in the fixed-order process
  and are produced by the PS.} gives better stability against theory
uncertainties such as renormalisation and factorisation scale variations. But
this is a lesser benefit, and perhaps not important for many use-cases: a
realistic appreciation of how SHG MC will be \emph{used} at the HL-LHC is
crucial in the calculus of estimating physics demands, rather than assuming that
``more $n$s'' is always the best strategy.

Even less technical, but potentially crucial, are ``social'' initiatives, such
as better coordination between experiments. A great deal of MC generation CPU is
effectively duplicated, with ATLAS and CMS generating essentially equivalent MC
events. Little or no experiment intellectual property resides in these
pre-detector MC samples, and so the science case for keeping them private is not
strong. The evolution of data archival and distribution machinery through CERN,
cf.~the EOS, CVMFS, and Zenodo
systems~\cite{mascetti2015cernbox,peters2011exabyte,buncic2010cernvm,potter2015making},
means that sharing of events, at least those expensively generated at parton
level, is a very low-hanging fruit indeed for improving each experiment's ratio
of physics impact to CPU expenditure. While na\"ive suggestions of a factor of
two gain here are overblown (as mentioned, ATLAS and CMS have quite large
differences in generator preference), the simple availability of more variety
can only be a good thing. It is unlikely that such sharing would significantly
reduce any cross-checking via MC variation, since a greater variety of inputs
intrinsically reduced biases, and the availability of common pre-detector
samples should assist comparison of the experiments' measurements in slightly
different phase-spaces.  The development of the new HDF5 HPC partonic-event
format may end up being a major player in this, along with the required
engineering work to factorise previously monolithic generators into re-runnable
processing stages. Even such a conceptually easy plan requires real effort to
turn it into reality, but efforts coordinated by the HSF suggest that openness
and sharing of MC-generated data is now on the agenda to an unprecedented
extent.

Lurking in the background is also a big question of incentives and institutional
structures, raised in policy statements by the HSF~\cite{Alves:2017she} and
MCnet~\cite{Buckley:2019kjt}. Particle physics internationally has a dichotomy
of funding and career streams for ``experimental'' vs. ``theoretical''
personnel, and SHG MC development sits uncomfortably on the divide -- essential
to experiments without itself being experimental physics, but containing a
significant amount of engineering and data-comparison work that is not always
valued by theory panels.  SHG MC authors operate in the theory environment and
respond to that community's pressures -- publishing of identifiably theoretical
papers, and an emphasis on formal innovation over computational
performance. This has been characterised~\cite{anontheorist} as an incentive to
make MC codes that run just fast enough to write the theory paper, but no
faster! This is all understandable, if not optimal, but means that if LHC
experiments want improvements in MC generation performance -- as is all but
mandated by the HL-LHC physics demands and computing budget -- they will need to
contribute directly to the technical development and performance optimisation of
MC codes. This includes ``community codes'' like the LHAPDF parton density and
the HepMC event record libraries~\cite{Buckley:2014ana,Dobbs:2001ck}, neither of
which has dedicated funding: significant CPU is wasted in these ``trivial''
codes, for lack of developer person-power. Discussions along these lines are in
an early stage and need to be taken seriously: to get the best out of our
massive experimental facilities, a collegiate and two-way relationship is needed
between the experiments and the MC theorists whose work we rely on.



\section{Conclusions}

This has been a whistle-stop tour through the current state of fully-exclusive
MC event generation, primarily from the perspective of current and future LHC
experiment simulation campaigns, and their increasingly tight CPU budgets. As we
have seen, the main computational costs of MC generation are the efficient
sampling and merging of high-multiplicity partonic matrix elements, a problem
which becomes exponentially worse when the additional divergence structures of
loop-level matrix elements are introduced. Several promising strategies to
ameliorate this problem have been shown: use of machine learning both for
generic and process-specific improvements to ME samplers, and factorisation of
the ME calculation part for efficient use of MPI HPC facilities. In addition we
have seen that ML techniques may be used for non-parametric estimation of
calculation elements currently beyond \emph{a priori} calculation. It is
important also to not lose sight of the physics in such calculations and how it
fits into the experiments' usage patterns: more collaboration, both between the
experiments and between experiments and MC authors, and judicious use of ``less
precise'' MC samples have potential to return dividends as significant as the
technical acrobatics.  As the LHC Run~3 and HL eras approach, we may hope to see
more of all these approaches in experiment production campaigns.


\section*{Acknowledgements}

Many thanks to Stefan Hoeche, Holger Schulz, Keith Hamilton, Marek Schoenherr,
Frank Siegert, Josh McFayden, Valentin Hirschi, Andrea Valassi and more,
particularly in the MCnet and HSF communities. This work has received funding
from the European Union's Horizon 2020 research and innovation programme as part
of the Marie Sklodowska-Curie Innovative Training Network MCnetITN3 (grant
agreement no.~722104), and the Royal Society via University Research Fellowship
grant no~UF160548.

\section*{References}
\bibliographystyle{unsrt}
\bibliography{acat-mc-plenary}

\end{document}